\documentstyle[prl,aps,epsfig,prbbib]{revtex}
\begin{document}
\draft
\twocolumn[\hsize\textwidth\columnwidth\hsize\csname  
@twocolumnfalse\endcsname

\title{Pairing fluctuation effects on the single-particle spectra for the 
superconducting state}
\author{P. Pieri, L. Pisani, and G.C. Strinati}
\address{Dipartimento di Fisica, UdR INFM,
 Universit\`{a} di Camerino, I-62032 Camerino, Italy}
\date{\today}
\maketitle
\hspace*{-0.25ex}

\begin{abstract}
Single-particle spectra are calculated in the superconducting state for a
fermionic system with an
attractive interaction, as functions of temperature and coupling
strength from weak to strong.
The fermionic system is described by a single-particle self-energy that
includes pairing-fluctuation
effects in the superconducting state.
The theory reduces to the ordinary BCS approximation in weak coupling and
to the Bogoliubov approximation
for the composite bosons in strong coupling.
Several features of the single-particle spectral function are
shown to compare favorably with
experimental data for cuprate superconductors.
\end{abstract}
\pacs{PACS numbers: 03.75.Ss, 03.75.Hh, 05.30.Jp}
\hspace*{-0.25ex}
]
\narrowtext

Information on the single-particle spectral function, that is obtained
from ARPES~\cite{damascelli} and tunneling data~\cite{kugler} for cuprate 
superconductors, can shed light
on the characteristic features of the superconducting state as well as on its 
connection 
with the unconventional normal state above the critical temperature $T_{c}$.

Most prominent among these features are the continuous evolution of a broad
pseudogap structure from above to below $T_{c}$,~\cite{damascelli,kugler} the 
emergence of a coherent peak below $T_{c}$ that combines with
the pseudogap structure to yield a characteristic peak-dip-hump 
profile~\cite{damascelli} (for which two distinct energy scales can be 
identified), and the peculiar dependence of the frequency
position and weight of the coherent peak on temperature and 
doping.~\cite{damascelli}
Generally speaking, features of the standard BCS theory are recovered for
overdoped samples, while non-BCS behaviors occur for optimally-doped  and
underdoped samples.

The origin of the peak-dip-hump profile has especially been the subject of
controversy, being attributed either to ``extrinsic'' effects like the 
bilayer splitting~\cite{bilayer} or to ``intrinsic''effects.
The latter can be identified over and above the extrinsic effects, and are 
believed to originate from strong (many-body) interactions in the system
~\cite{intrinsic}.

Quite independently from the microscopic origin of the fermionic attraction
giving rise to superconductivity,
its strength is believed to be stronger in optimally-doped and
underdoped than in overdoped samples~\cite{ong}, consistently with the above 
findings.
This implies that pairing fluctuations should definitely be taken into
account, irrespective of
other effects (such as the bilayer splitting and/or additional
many-body effects associated with specific pairing mechanisms~\cite{norman}).

In this paper, we assess the role of pairing fluctuations for the
single-particle spectral function in
the superconducting state on rather general grounds, by identifying a
single-particle self-energy that describes
fluctuating Cooper pairs in weak coupling and non-condensed composite
bosons in strong coupling.
(The latter form as bound-fermion pairs in the strong-coupling limit of the
fermionic attraction.)
To this end, we consider fermions mutually interacting via an attractive 
contact potential in a 3D continuum,
without taking into account lattice effects nor the explicit physical mechanism
responsible for the attraction.
Qualitative comparison with experimental data will thus rest on identifying
the increasing coupling
strength in this model with the increasing potential strength for decreasing
doping level in the phase diagram of cuprate superconductors.
Although the effective model here considered is 
oversimplified for a full
description of cuprates, the physical questions we are addressing are 
sufficiently general that
this minimal model will prove sufficient to capture the main experimental 
features.

Our main results for the single-particle spectral function (to be compared
with the experimental findings)
are summarized as follows:

\noindent
(i)   A broad pseudogap structure and a coherent peak are simultaneously
present in a wide
      range of coupling and temperature below $T_{c}$, giving rise to a
peak-dip-hump profile.
Two distinct energy scales (the positions 
$\Delta_{{\rm pg}}$ of the broad pseudogap structure and
$\Delta_{\rm{m}}$ of the coherent peak) can then be
extracted from the spectra at given temperature and coupling, as seen
experimentally~\cite{damascelli}.

\noindent
(ii)  At a given (notably, intermediate) coupling, BCS-like
features coexist with non-BCS behaviors.
The position $\Delta_{\rm{m}}$ and weight $z$ of the coherent peak
versus wave vector follow a BCS-like dependence, as found 
experimentally~\cite{Campuzano96,Sato03}.
At the same time, the weight $z$ of the coherent peak has a strong
temperature dependence, in agreement with experiments~\cite{Feng,Ding} but 
in contrast with BCS theory.

\noindent
(iii) At low temperature, the weight $z$ of the coherent peak has a strong
dependence on coupling, decreasing monotonically from weak to strong 
coupling (as seen experimentally in cuprates for decreasing 
doping~\cite{Feng,Ding}).
At the same time, the position $\Delta_{\rm{m}}$ of the coherent peak
increases monotonically with coupling~\cite{Feng,Ding}.

\noindent
(iv)  The positions $\Delta_{{\rm pg}}$ of the broad pseudogap structure at
$T_{c}$ and $\Delta_{\rm{m}}$ of the coherent peak near zero temperature 
cross each other as function of coupling about intermediate coupling.
This feature is also seen experimentally by intrinsic tunneling 
experiments at different dopings~\cite{krasnov} (although these data are
subject to controversy\cite{zav03}).

Pairing fluctuations are taken into account in our theory by considering, 
besides the off-diagonal BCS-like
self-energy, the diagonal $t$-matrix self-energy suitably extended to the
superconducting state
(we use the Nambu formalism throughout).
The $t$-matrix self-energy has been widely used to include pairing
fluctuations in the normal state~\cite{NSR}, and specifically to account for 
pseudogap features in the single-particle spectral
function~\cite{PPSC-02}. In our theory,
the diagonal $t$-matrix self-energy (that survives above $T_{c}$) will 
essentially be responsible
for the presence of the broad pseudogap structure below $T_{c}$. The 
off-diagonal BCS-like self-energy will instead give rise to the simultaneous 
emergence of the coherent peak.
In addition, our theory recovers the 
Bogoliubov approximation for the composite bosons in strong 
coupling.~\cite{leo03} 

The diagonal $t$-matrix self-energy reads
\begin{equation}
\Sigma_{11}(k)= - \int\frac{d {\bf q}}{(2\pi)^{3}} \frac{1}{\beta} 
\sum_{\Omega_{\nu}}\Gamma_{11}(q){\mathcal G}_{11}(q-k)\; , 
\end{equation}
while the off-diagonal self-energy has the BCS-like form 
$\Sigma_{12}(k)= -\Delta$ in terms of the superconducting order
parameter $\Delta$.
The pairing-fluctuation propagator $\Gamma_{11}$ in the 
broken-symmetry phase entering Eq.~(1) is given by:
\begin{equation}
\Gamma_{11}(q) = \frac{\chi_{11}(-q)}{\chi_{11}(q) \chi_{11}(-q) - 
\chi_{12}(q)^{2}} 
\label{gamma11}
\end{equation}
with
\begin{eqnarray}
- \chi_{11}(q) &=& \frac{m}{4\pi a_F} +  \int \! \frac{d {\mathbf p}}{(2\pi)^{3}} \left[
\frac{1}{\beta} \sum_{\omega_n} 
{\mathcal G}_{11}(p+q) {\mathcal G}_{11}(-p) 
\right.\nonumber\\
& &\phantom{\frac{m}{4\pi a_F}} - 
\left.\frac{m}{{\bf p}^2}\right]\label{A-definition}\\
\chi_{12}(q) & = & \int \! \frac{d {\mathbf p}}{(2\pi)^{3}} 
\frac{1}{\beta} \sum_{\omega_n} 
{\mathcal G}_{12}(p+q) {\mathcal G}_{21}(-p)\; .
\label{B-definition}
\end{eqnarray}
In the above expressions, $q=({\mathbf q},\Omega_{\nu})$, 
$k=({\mathbf k},\omega_{l})$, and $p=({\mathbf p},\omega_{n})$ (${\mathbf q}$, 
${\mathbf k}$, and ${\mathbf p}$ being wave vectors,
$\Omega_{\nu}$ a bosonic Matsubara frequency, and $\omega_{l}$ and 
$\omega_n$ fermionic Matsubara frequencies),
$m$ is the fermionic mass, $\beta$ the inverse
temperature, and ${\mathcal G}_{ij}$  $(i,j=1,2)$ have the form of the BCS
single-particle Green's functions
in Nambu notation.

For given temperature and strength of the point-contact interaction, the
chemical potential $\mu$
and the order parameter $\Delta$ are obtained by solving the coupled
particle number and gap
equations:
\begin{eqnarray}
n & = &  2 \, \int \! \frac{d {\mathbf k}}{(2\pi)^{3}} \, \frac{1}{\beta}
\sum_{\omega_l} \, e^{i\omega_{l}\eta}
            \, G_{11}({\mathbf k},\omega_{l})   
\label{n-G-11}\\
-\frac{m}{4 \pi a_{F}} & = & \int \! \frac{d{\mathbf k}}{(2\pi)^{3}} \, \left[
\frac{\tanh(\beta E({\mathbf k})/2)}{2 E({\mathbf k})} \, - \,
\frac{m}{{\mathbf k}^{2}} \right] 
\label{BCS-gap_equation}
\end{eqnarray}
where $\eta=0^+$ and $E({\mathbf k})=[\xi({\mathbf k})^2+\Delta^2]^{1/2}$ 
with $\xi({\mathbf k})={\mathbf k}^2/(2 m)-\mu$.
The scattering length $a_{F}$ of the (fermionic) two-body problem has been
introduced to
eliminate the ultraviolet divergences originating from the use of a
point-contact interaction.
The dimensionless parameter $(k_{F}a_{F})^{-1}$ (where $k_{F}$ is the Fermi
wave vector related to the density $n$ via $k_{F}= (3\pi^2 n)^{1/3}$) 
characterizes the fermionic coupling strength and ranges
formally from $- \infty$ to $+ \infty$.
[Correspondingly, the Fermi energy $\varepsilon_{F}$ equals $k_{F}^{2}/(2m)$.]
In practice, the crossover from the weak- to strong-coupling regimes
occurs in the limited range
$ -1 \lesssim (k_{F}a_{F})^{-1} \lesssim +1$, which will be mostly explored
in what follows.
The dressed normal (diagonal) Green's function 
$G_{11}$ in 
Eq.~(\ref{n-G-11}) is obtained by matrix inversion of the 
Dyson's equation in Nambu formalism with the above self-energies and is given 
by:
\begin{eqnarray}
G_{11}({\mathbf k},\omega_{l}) &=& \left[\,
i\omega_{l}-\xi({\mathbf k})-\Sigma_{11}({\mathbf k},\omega_{l})
\phantom{\frac{1^2}{1}}\right.\nonumber\\
&-&\left.
\frac{\Delta^2}{i\omega_{l}  +  \xi({\mathbf k})
+ \Sigma_{11}({\mathbf k},-\omega_{l})}\right]^{-1}\; .
\label{G-11-Matsubara} 
\end{eqnarray}
Note that, while the number equation (\ref{n-G-11}) contains the dressed 
normal (diagonal) Green's function, the gap equation (\ref{BCS-gap_equation}) 
is obtained from the anomalous (off-diagonal) BCS Green's function and thus 
its form is not modified with respect to the BCS theory.
This ensures that the pairing-fluctuation propagator (\ref{gamma11}) remains 
gapless for all temperatures (below $T_{c}$) and couplings.
[The value of $T_c$ is obtained from Eqs.~(\ref{n-G-11}) and
 (\ref{BCS-gap_equation}) by setting $\Delta=0$ identically.] The numerical 
values of the chemical potential $\mu$ and order parameter $\Delta$ at given 
temperature and coupling differ, however, from those obtained by BCS theory 
due to the different structure of the number equation.
It can be verified~\cite{leo03} from 
Eqs.~(\ref{gamma11})-(\ref{BCS-gap_equation}) that the Bogoliubov approximation
for the composite bosons is recovered in the strong-coupling 
($\beta \mu \rightarrow - \infty$ and $\Delta \ll |\mu|$) limit.
This represents a notable achievement of our theory.

The use of bare BCS single-particle Green's functions in the
self-energy (1) further enables us to
perform the analytic continuation to real frequency $\omega$ in a closed
form, as to avoid numerical extrapolation procedures.
To this end, and similarly to what was done in Ref.~\onlinecite{PPSC-02} for 
the normal phase, we express the pairing-fluctuation propagator 
$\Gamma_{11}$ by its spectral representation:
\begin{equation}
\Gamma_{11}({\mathbf q},\Omega_{\nu}) \, = - \frac{1}{\pi} 
\int_{-\infty}^{+\infty} \! d\omega' \,
\frac{{\rm Im} \Gamma_{11}({\mathbf q},\omega')}{i\Omega_{\nu} - \omega'}\; .
\label{spectral-representation}
\end{equation}
Here $\rm{Im}\Gamma_{11}({\bf q},\omega)$ is defined as the imaginary part of
$\Gamma_{11}({\bf q},i\Omega_{\nu} \rightarrow \omega + i\eta)$.
This quantity is, in turn, obtained from the expressions 
(\ref{gamma11})-(\ref{B-definition}), with the replacement
$i\Omega_{\nu} \rightarrow \omega + i\eta$ made \emph{after\/} the sum
over the internal (Matsubara)
frequency has been performed.
In this way, the imaginary part of the retarded self-energy
$\Sigma^{R}_{11}$ is:
\begin{eqnarray}
{\rm Im}\Sigma_{11}^R({\bf k},\omega)&=&-\int \frac{d {\bf q}}{(2\pi)^3}
\left\{u^2_{{\bf q}-{\bf k}} 
{\rm Im}\Gamma_{11}({\bf q},\omega+E({\bf q}-{\bf k}))\right.\nonumber\\ 
&\times& \left[f(E({\bf q}-{\bf k}))+b(\omega+E({\bf q}-{\bf k}))\right] 
\nonumber 
\\
&+ & v^2_{{\bf q}-{\bf k}} {\rm Im}\Gamma_{11}({\bf q},
\omega-E({\bf q}-{\bf k}))\nonumber\\
 &\times& \left.\left[f(-E({\bf q}-{\bf k}))+b(\omega-E({\bf q}-{\bf k}))
\right] \right\}. 
\label{imsig11}
\end{eqnarray}
Here, $f(x)=[\exp(\beta x)+1]^{-1}$ and $b(x)=[\exp(\beta x)-1]^{-1}$ are the
Fermi and Bose distributions, while 
$u^2_{{\bf k}}$ and $v^2_{{\bf k}}$ are the usual BCS coherence factors.
The real part of $\Sigma^{R}_{11}$ is then obtained via a
Kramers-Kronig transform.
The retarded single-particle Green's function $G^{R}_{11}$
results from these functions,
in terms of which the single-particle spectral function 
$
A({\bf k},\omega) =-\frac{1}{\pi} {\rm Im} G_{11}^R({\bf k},\omega)
$
of interest is eventually obtained as a function of $\omega$ for any given 
${\bf k}$.
\cite{footnote}

\begin{figure}[h]
\centerline{\epsfig{figure=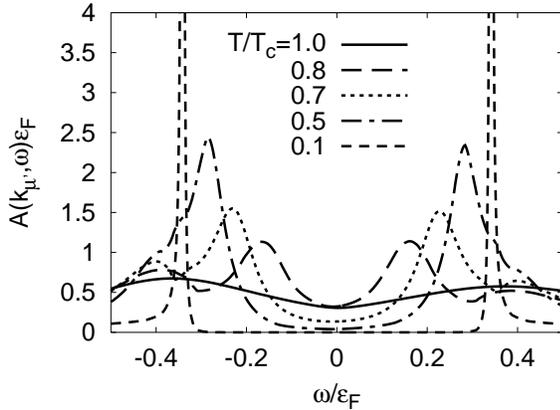,width=8.1cm}}
\caption{Single-particle spectral function vs frequency (in units
of $\varepsilon_{F}$) at different temperatures, for $|{\bf k}| = k_{\mu'}$ 
and $(k_{F}a_{F})^{-1} = -0.45$.}
\label{spectT}
\end{figure}

Figure~1 shows the evolution with temperature of the single-particle
spectral function for an intermediate
coupling ($(k_{F}a_{F})^{-1} = -0.45$). [For this coupling,
the ratio of the pair-breaking temperature $T^*$ to the  
critical temperature $T_c$ is about 1.25, as obtained in 
Ref.~\onlinecite{PPSC-02}.]
Focusing specifically on the features at negative $\omega$, note how the 
(sharp) coherent peak grows from the (broad) pseudogap structure already 
present at $T_{c}$.
The coherent peak becomes sharper upon lowering the temperature and gains
weigth at the expenses of the
pseudogap structure, giving rise to a characteristic peak-dip-hump profile.
The two features coexist over a wide range of temperature.

Figure~2 shows the wave-vector dependence of the single-particle
spectral function at the
temperature $T= 0.7 T_{c}$ for the same coupling of Fig.~1.
Following the evolution of the coherent peak across the underlying Fermi
surface, one identifies the characteristic particle-hole 
mixing of the BCS theory, with a reflection of both particle and hole bands
accompanied by a transfer of the spectral weight from negative to positive
frequencies.
In our theory, such BCS-like features coexist with non-BCS behaviors, namely,
the occurrence of the pseudogap structure and the strong temperature 
dependence of the spectral weight of the coherent peak (see Fig.~4 below).

\begin{figure}[h]
\centerline{\epsfig{figure=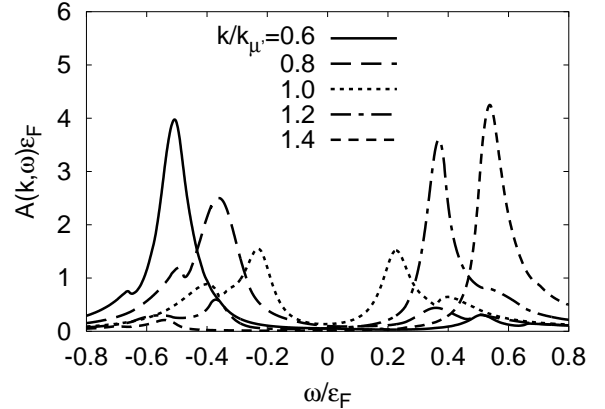,width=8.1cm}}
\caption{Single-particle spectral function vs frequency (in units
of $\varepsilon_{F}$) at different wave vectors about $k_{\mu'}$, for 
$T= 0.7 T_{c}$ and $(k_{F} a_{F})^{-1} = -0.45$.}
\label{momentumevol}
\end{figure}
\vspace{-2truecm}
\begin{figure}[h]
\centerline{\epsfig{figure=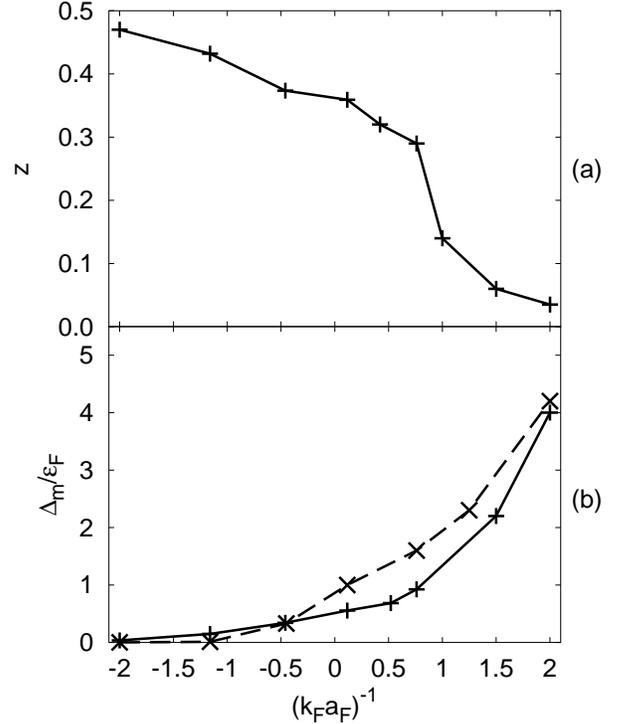,width=8.1cm}}
\vspace{-2.4truecm}
\caption{(a) Weight $z$ and (b) position $\Delta_{{\rm m}}$ (full line) of the
coherent peak at negative frequencies vs coupling for $T= 0.1 T_{c}$.
In (b) the pseudogap $\Delta_{{\rm pg}}$ for $T=T_c$ (dashed line) is also 
shown.}
\label{coupldep}
\end{figure}

Figures~3(a) and 3(b) report, respectively, the weight $z$ and position
$\Delta_{{\rm m}}$ of the coherent
peak of the single-particle spectral function at negative frequencies vs
coupling for a low temperature
($T= 0.1 T_{c}$).
The peak weight $z$ saturates at the BCS value $0.5$ in weak coupling,
decreases markedly across the crossover region 
$ -1 \lesssim (k_{F}a_{F})^{-1} \lesssim +1$, and becomes negligible
in strong coupling.
At the same time, the peak position $\Delta_{\rm m}$ increases 
monotonically across the crossover region.
In Fig.~3(b) we also report the pseudogap $\Delta_{{\rm pg}}$ (dashed line),
as identified from the position of the maximum of the spectral function at 
$T_{c}$.
While the qualitative trend of $\Delta_{{\rm m}}$ and
$\Delta_{{\rm pg}}$ vs coupling is similar,
the two curves cross each other at about the intermediate-coupling value
$(k_{F}a_{F})^{-1} = -0.45$. In addition, we have verified that 
$\Delta_{\rm m}$ about coincides with the value of the order parameter $\Delta$
in the weak-to-intermediate coupling region. 

\begin{figure}[h]
\centerline{\epsfig{figure=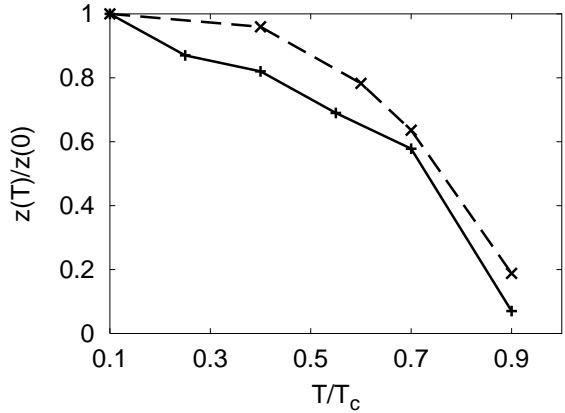,width=8.1cm}}
\vspace{.25truecm}
\caption{Temperature dependence of the weight $z$ of the coherent
peak at negative frequencies (full line) for $(k_{F} a_{F})^{-1} = -0.45$. 
The superfluid density $\rho_{s}$ (dashed line) is also shown for comparison.}
\label{weightdep}
\end{figure}

Figure~4 reports the dependence of the weight $z$ on temperature
for the coupling value
$(k_{F}a_{F})^{-1} = -0.45$ (full line).
Note the strong temperature dependence of this quantity, which vanishes at
$T_{c}$. This contrasts the BCS behavior, whereby the weight of the coherent 
peak for negative
frequencies would equal $0.5$
irrespective of temperature.
The temperature dependence of the superfluid density $\rho_{s}$ (calculated
according to Ref.~\onlinecite{aps03})
is also shown in the figure (dashed line).
The resemblance between these two quantities has been noted
experimentally for nearly-optimally-doped cuprates~\cite{Feng,Ding}.
We have, however, verified that this resemblance does not occur for
other values of the
coupling, both on the weak- and strong-coupling sides of the crossover.
On the weak-coupling side of the crossover, in particular, our finding is 
confirmed by the BCS theory, whereby $z=0.5$
irrespective of temperature while $\rho_{s}$ decreases monotonically from
$T=0$ to $T_{c}$.
For this reason, no universal correspondence between the temperature
dependence of $z$ and $\rho_{s}$
should be expected on physical grounds.

The results shown in the above figures refer mostly to the
weak-to-intermediate coupling side of the
crossover region.
Our conclusions have, however, been drawn from a sistematic study 
of the whole crossover region from weak to strong coupling, of which Fig.~3 
is an example.

All the qualitative features extracted from the above figures compare favorably
with the experimental data
on cuprates, as anticipated in the points (i)-(iv).
This qualitative comparison rests on the assumed correspondence between the
increasing of the coupling
strength in the present theory and the decreasing doping level in the phase 
diagram of cuprate superconductors. 

In conclusion, we have shown that pairing fluctuations 
can (at least qualitatively) account for several nontrivial features of 
single-particle spectra in the superconducting
state.
Our results specifically demonstrate that the experimental finding of two
distinct features (pseudogap
structure and coherent peak) in the single-particle spectral function is
fully consistent with the occurrence
of strong pairing fluctuations in cuprate superconductors.
Competition of two distinct order 
parameters is therefore not required to account for the occurrence of two 
different energy scales in the experimental data.

We are indebted to A. Perali for discussions. Financial support from the
Italian MIUR under contract COFIN 2001
Prot. 2001023848 is gratefully acknowledged.

\end{document}